# Closed-loop machine learning for discovery of novel superconductors

Elizabeth A. Pogue[a,1,2], Alexander New[a,1], Kyle McElroy[a], Nam Q. Le[a], Michael J. Pekala[a], Ian McCue[e], Eddie Gienger[a], Janna Domenico[a], Elizabeth Hedrick[b], Tyrel M. McQueen[b,c,d,2], Brandon Wilfong[c,d], Christine D. Piatko[a], Christopher R. Ratto[a], Andrew Lennon[a], Christine Chung[a], Timothy Montalbano[a], Gregory Bassen[c,d], and Christopher D. Stiles[a,2]

[a]Research and Exploratory Development Department, Johns Hopkins University Applied Physics Laboratory, 11100 Johns Hopkins Road, Laurel, 20723, Maryland, United States of America; [b]Department of Materials Science and Engineering, Johns Hopkins University, 3400 N. Charles Street, Baltimore, 21218, Maryland, United States of America; [c]Department of Chemistry, Johns Hopkins University, 3400 N. Charles Street, Baltimore, 21218, Maryland, United States of America; [d]Institute for Quantum Matter, William H. Miller III Department of Physics and Astronomy, Johns Hopkins University, 3400 N. Charles Street, Baltimore, 21218, Maryland, United States of America; [e]Department of Materials Science and Engineering, Northwestern University, 2220 Campus Drive, Evanston, 60208, Illinois, United States of America

This manuscript was compiled on December 21, 2022

**The discovery of novel materials drives industrial innovation ([1]), although the pace of discovery tends to be slow due to the infrequency of "Eureka!" moments ([2]). These moments are typically tangential to the original target of the experimental work: "accidental discoveries". Here we demonstrate the acceleration of intentional materials discovery – targeting material properties of interest while generalizing the search to a large materials space with machine learning (ML) methods. We demonstrate a closed-loop ML discovery process targeting novel superconducting materials, which have industrial applications ranging from quantum computing to sensors to power delivery ([3]–[6]). By closing the loop, i.e. by experimentally testing the results of the ML-generated superconductivity predictions and feeding data back into the ML model to refine, we demonstrate that success rates for superconductor discovery can be more than doubled ([7]). In four closed-loop cycles, we discovered a new superconductor in the Zr-In-Ni system, re-discovered five superconductors unknown in the training datasets, and identified two additional phase diagrams of interest for new superconducting materials. Our work demonstrates the critical role experimental feedback provides in ML-driven discovery, and provides definite evidence that such technologies can accelerate discovery even in the absence of knowledge of the underlying physics.**

Closed-loop machine learning | Superconductivity | Materials discovery

The discovery of new materials with quantum properties, like superconductivity, drives industrial and scientific advancement ([8]). However, the difficulty in creating novel superconductors from potential compositions has made the discovery process stochastic and unpredictable ([9]). For the past quarter century, statistical approaches have aimed to better understand and predict superconductivity ([10]), with Machine learning (ML) emerging as the preferred approach ([11]–[15]). Although ML has shown some success in predicting novel superconductors that do not exist in known databases ([13]–[15]), many remain untested because it is not clear that many of these predictions are promising ([11], [12]). ML models also commonly make predictions that clearly will not superconduct for reasons discussed in this paper. Previous studies have had limited success because they fail to consider the end-to-end materials discovery lifecycle. They treat materials and databases of material properties as fixed snapshots rather than evolving systems. Additionally, they neither incorporate physics nor account for the biases of experimental sciences in the training of ML models. Once a new superconductor has been identified, experimentalists generally search for superconductivity in related materials, accelerating the discovery of new superconductors but biasing the datasets used for machine learning to favor certain families of compounds.

Here we report on combining ML techniques with materials science and physics expertise to "close the loop" of materials discovery (Figure 1). We demonstrate that previous ML models fail to generalize across diverse materials spaces (see SI), making them unlikely to identify superconductors that are dissimilar to known ones. Consequently, we alternate between ML property prediction and experimental verification to continuously improve the fidelity of ML property prediction in regimes sparsely represented by existing materials databases. Crucially, this adds both negative data (materials incorrectly predicted to be superconductors) and positive data (materials correctly predicted) to ML training, enabling the ML model's overall representation of the space of materials to be iteratively refined. This also avoids assuming that some materials are non-superconductors without verification ([13]). "Closing the loop"—leveraging inputs from materials experts and experimental verification of predictions—allowed us to demonstrate the first ML-guided discovery of a novel superconductor.

Our process uses active learning ([16]) to iteratively select data points to be added to a training set, with the overall goal of identifying materials with desirable properties. After ML model prediction, we use human domain expertise to further filter predicted candidates to those deemed most likely to superconduct. We engaged in a small number of total prediction/experimental measurement iterations; to maximize the superconducting transition temperatures ($T_c$s) of superconductors discovered over further iterations, we can use acquisition functions developed for Bayesian Optimization ([17], [18]). Our approach retains a human-in-the-loop for synthesizing and characterizing materials, but further automation is possible, involving, e.g., ML systems selecting experiments to be con-ducted, or robot-powered self-driving laboratories (19–21).

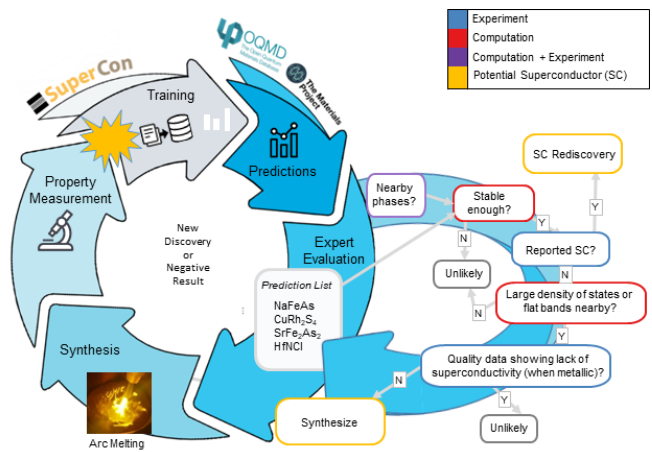

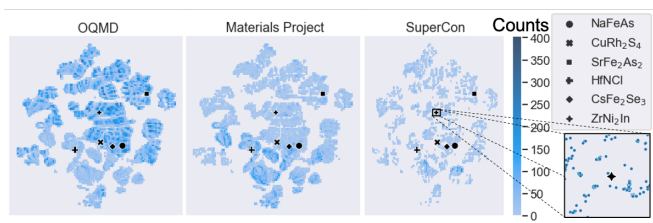

**Fig. 1. ML Prediction-expert evaluation-experimental measurement loop and expert evaluation process.** We closed the ML prediction-experimental measurement loop four times in this study. To improve success, experts evaluated lists of predicted superconductors following the procedure outlined to the right to determine which predictions were most promising. When evaluating these predictions, the data sources referenced are denoted by color. "Experiment" includes both literature data and prior but unpublished internal data. "Computation" includes data from computational databases, computed literature data, and targeted computational data that we acquired.

**Fig. 2. Visualization of OQMD, Materials Project and SuperCon Databases together.** A histogram of the concentration of materials from a Uniform Manifold and Projection (UMAP) (25) embedding of OQMD (without superconductivity information), MP (without superconductivity information), and SuperCon (superconductivity information), based on Magpie (26) descriptors for the datasets. The embedding is learned from concatenation of Magpie descriptors obtained from all three datasets; the same axis limits are used across each subplot. $T_c$ is not part of the Magpie descriptors and, therefore, did not influence the representation. The six black symbols indicate five rediscovered superconductors (Section 4), and our novel superconductor, near $ZrNi_2In$. The inset on the right highlights the local region in which $ZrNi_2In$ is found, which is sparse and far from the rediscovered superconductors.

We confirmed our approach by rediscovering five superconductors outside of the ML model's training set. Through closing the prediction-experimental measurement loop four times, we discovered a new superconductor in the Zr-In-Ni phase diagram, and identified two other phase diagrams of interest (Zr-In-Cu and Zr-Fe-Sn). These materials come from a wide variety of families: iron pnictides, doped 2D ternary transition metal nitride halides, intermetallics, and spinels.

## 1. ML Models for predicting superconductivity

For the initial prediction step of the closed-loop approach, we trained an ML model to predict the superconducting transition temperature, $T_c$, of candidate materials. Our primary source of training data, SuperCon (22), contains compositions of known superconductors. Only the materials' compositions were used to train the ML model for predicting $T_c$ since SuperCon did not contain additional usable information. Materials Project (MP) (23) and Open Quantum Materials Database (OQMD) (24), some of the largest public sets of computational materials data, supplied candidate compositions to be screened for superconductivity. These two databases do not contain any $T_c$ data. These three datasets are visualized in Figure 2 using a joint representation. Crucially, the amount of data for which we have superconducting information is much smaller than our other sources of data and is not uniformly sampled across the joint space.

It is well-known (27) that when ML methods make predictions on data outside of their training data distribution, accuracy often suffers; this is often called the *out-of-distribution generalization problem*. In cheminformatics (28), it is common to assess whether a dataset is within the distribution of a training dataset by seeing how far, in some representative metric space, its points are from the training dataset: as the difference between the distribution of new data and the training data increases, the likelihood that a model will accurately predict their properties decreases. To improve assessment of generalization, out-of-distribution data may be simulated by creating validation sets that split based on non-random criteria like Murcko scaffold (29) or cluster identity, the latter being the leave-one-cluster-out cross-validation (LOCO-CV) strategy (30). In the Supplementary Information and Extended Data, we apply LOCO-CV in a simulated superconductor-identification problem. We show that a strong ML model fails to make accurate predictions of superconducting status on out-of-distribution data. This motivates our need for multiple iterations of model training, candidate selection, candidate synthesis, and model retraining.

We rely on a recent ML model for chemical property prediction, Representation learning from Stoichiometry (RooSt) (see Methods and SI), to predict a material's superconductivity using only its stoichiometry (i.e., ignoring the material's crystal structure). This was necessary because the SuperCon database used for training contains insufficient structural information for more expressive ML models, such as the crystal graph convolutional neural network (31, 32). The intersection of SuperCon with MP and OQMD was not large enough to allow us to easily estimate the corresponding structure.

**A. Identifying new predicted superconductors using ML.** Following the training of an ensemble of RooSt models using the SuperCon database, we apply them to our set of potential superconductors (i.e., MP and OQMD). We filter for materials likely to be high-$T_c$ superconductors (see Methods, Extended Data, and SI).

A risk of searching for new superconductors from a static list of candidates is that while a material in MP or OQMD may not have the exact composition as a superconductor, it may have a composition extremely close in terms of stoichiometry, such as $MgB_2$ vs. $Mg_{33}B_{67}$. Thus, every time we produce a new list of candidates, we identify each candidate's minimal Euclidean distance, in Magpie-space (26), to any point in our training data, and we remove candidates too close to SuperCon. This distance is used by domain experts as one of the metrics to select final candidates to synthesize and test.

## 2. Incorporating domain expertise

It is not practical to experimentally verify all ML predictions. The costs associated with fabricating and characterizing a new material are high; hence we are only able to experimentally analyze a small subset of the ML predictions. Expert analysis to factor in information not present in the ML training data (Figure 1) was key for achieving high success rates.

The Materials Project and OQMD databases both contain calculated stability information not used by the ML model. Of 190 predicted superconductors in a given prediction round, only 39 compounds were calculated to be stable ($E_{over\ hull} = 0.00\,\text{eV/atom}$) but 83 were nearly stable ($E_{over\ hull} < 0.05\,\text{eV/atom}$). Stable materials and those with prior experimental reports were prioritized to increase the likelihood that targeted compounds could be successfully synthesized. Prioritizing these materials ensured that failures to observe superconductivity were indicative of the behavior of the targeted compound rather than a failure to synthesize that compound.

Insulating materials like $\beta$-ZrNCl and the cuprates superconduct with high $T_c$s because they can be doped into a metallic state (33). One long-running challenge for machine-learning approaches to predicting high-$T_c$ superconductivity is that large bandgap insulators incapable of superconductivity tend to be given overweighted classification scores, likely due to the high $T_c$s of the cuprates(15). Therefore, metals and easily doped materials were favored for testing. Similarly, for some predicted metals, we investigated nearby compounds with similar structures that were known in literature but were not found in Materials Project or OQMD (e.g.: $Zr_3Fe_4Sn_4$ and $Hf_3Fe_4Sn_4$ (34, 35)) and isostructural compounds with promising band structures (e.g.: $ZrNi_2In$).

Since the $T_c$s of compounds are very sensitive to alloy disorder and lattice parameter, we explored several compositions near each prediction (36). We also considered the ease and safety of synthesizing the target materials (e.g., by excluding extremely high-pressure syntheses). Powder X-ray diffraction (XRD) was used to ensure that the target material was successfully made and temperature-dependent AC magnetic susceptibility was used to screen for superconductivity. Superconductors are perfectly diamagnetic below their $T_c$ with no (or minimal) applied field.

## 3. High-throughput experimental evaluation of $T_c$ using A15 phases

To illustrate the sensitivity of experimentally-measured $T_c$s to processing conditions, we made and tested samples with $A_3B$ stoichiometry (Figure 3a), including many known superconductors from the A15 family (44). Similar compositional sensitivity is common in other systems beyond A15 compounds. For example, as $x$ varies between 0 and 0.35, $La_{2-x}Sr_xCuO_4$ can vary from not superconducting to having a $T_c$ up to 36 K (14). Our experiments show that high-throughput synthesis and characterization techniques can reliably and quickly screen systems for superconductivity. Optimization of many superconducting phases requires much lower-throughput techniques for phase-pure and fully-superconducting samples.

## 4. Rediscovered Superconductors

Using this closed-loop method and high-throughput synthesis, we re-discovered five known superconductors that were not represented in the ML training dataset. A list of these is found in Table 1. Many of these superconduct when doped or at elevated pressures, illustrating the challenges of capturing doping effects and ruling out superconductivity purely on the basis of a negative result at atmospheric pressure.

We have experimentally sampled less than 100 different compounds and have been able to re-discover five previously reported (outside of the ML training set) superconductors and are investigating one novel superconductor and two new potential superconductors, demonstrating a much higher success rate than current exclusively human-lead approaches. A group of several teams of experts in Hosono et al. achieved a success rate of $\sim 3\,\%$ with 1000+ different compounds synthesized, in the wake of their discovery of iron-based superconductors in 2006 (7). Furthermore since, unlike conventional exclusively expert-driven investigations, our inputs are not limited to a single or few classes of materials, ML identifies potential families of candidate superconductor materials that may not have otherwise been explored.

**Table 1. Superconductors rediscovered by machine learning.** Since the model only took stoichiometry into account, predictions where small stoichiometry changes led to superconductivity were considered successful discoveries.

| Compound | Iteration | Database and Comments | $T_c$ |
|---|---|---|---|
| NaFeAs | 2 | Materials Project, reduced topotactically in water | 25 K (37) |
| $CuRh_2S_4$ | 2 | Materials Project | 4.7 K (38) |
| $SrFe_2As_2$ | 2 | Materials Project, reduced topotactically in water | 25 K (39) |
| HfNCl | 3 and 4 | Materials Project, OQMD, intercalated by lithium | 40 K (40) |
| $CsFe_2Se_3$ | 3 | Analogs $CsFe_{4-x}Se_4$ (5 K, $> \sim 34.55\,\text{GPa}$) (41), $BaFe_2Se_3$ (11 K, $> \sim 12.7\,\text{GPa}$) (42), and $BaFe_2S_3$ (24 K, $> \sim 10\,\text{GPa}$) (43) | |

## 5. New superconductor in Zr-In-Ni phase diagram

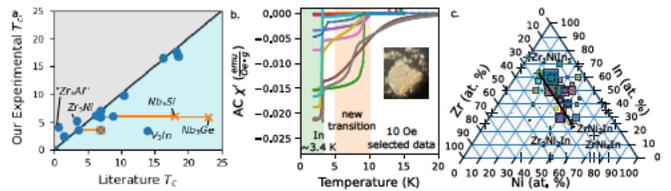

**Fig. 3. Experimental visualizations of superconductivity in A15 compounds and in the Zr-In-Ni system.** (a) Evaluation of our high-throughput synthesis of compounds with $A_3B$ stoichiometry (including A15 compounds) demonstrates the effects of processing on the measured $T_c$ and our ability to positively identify superconductors quickly. To track down the new superconductor in the Zr-In-Ni phase diagram, samples were tested from around the phase diagram (b and c) and the transition magnitude between 5 and 10 K (orange region) was used to scale data points. This transition was distinct from the indium-related transition (green). The strength of the indium signal could be used as a calibration to determine the relative strength of the higher-temperature signal. The line in (c) connects $Zr_2NiIn_5$ and $ZrNi_2In$. The compositions of samples with the strongest superconducting signals fell near this line. The colors in b, c, and Figure S5 and the symbols in b and Figure S5 match. The inset in (b) shows a piece of the sample with the strongest superconducting signal we observed.

Our novel ML-driven process discovered a new superconducting phase in the Zr-In-Ni system with a $T_c$ of $\sim 9\,\text{K}$ (Figure 1b-c and Extended Data). No other known elements, binaries or ternaries in the Zr-In-Ni system would explain a superconducting transition temperature this high and the elements and binaries have been extensively investigated (11, 44–46).

We synthesized many of these binaries and did not see evidence of superconductivity above 2 K unrelated to indium. A small, very broad diamagnetic signal was observed in a $ZrIn_3$ sample that went away upon annealing. The intensity of the superconducting signal was largest for indium and nickel-rich compositions connecting $ZrNi_2In$ and $Zr_2NiIn_5$ stoichiometries. The phases present in these arc melted samples matched well upon Rietveld refinement with $ZrNi_2In$, $Zr_2Ni_2In$, and In (See Extended Data, $R_{wp}$=1.25 %).

Magnetic susceptibility measurements for $ZrNi_2In$ reported in 1999 showed no evidence of a superconducting signal (47). However, in 2004, members of the same group presented at E-MRS that $ZrNi_2In$ superconducts with a $T_c$ of 9 K (48), suggesting there is variability in the system (There is no archival paper that we could locate and no data is readily available.) The isostructural compounds $ZrNi_2Ga$ and $ZrNi_2Al$ superconduct with a $T_c$ of 2.9 K and 1.38 K, respectively (49, 50). Their superconductivity has been attributed to a Van Hove singularity at the L point near the Fermi level that should also be present in $ZrNi_2In$ (23, 49). The atom contributing to these states is primarily nickel; moving down the periodic table to indium from gallium should not change these states significantly. Nearly phase-pure $ZrNi_2In$ had a weak signal compared to more In-rich compositions, indicating that the superconducting phase observed here is not due to stoichiometric $ZrNi_2In$.

The phase spaces near $ZrNi_2In$ and $ZrCu_2In$ would not have been identified as a place to look for higher-temperature superconductors, because Heusler alloys tend to have $T_c$'s well below 5 K. Our experimental investigation also revealed that the Zr-Cu-In and Zr-Fe-Sn systems merit further investigation (see Extended Data and SI).

## 6. Conclusions

We have presented the first ever closed-loop ML-based directed discovery of a novel superconductor with experimental verification (within the Zr-Ni-In system), identified two additional systems of interest (Zr-Cu-In and Zr-Fe-Sn), and rediscovered five others not represented in our ML training set.

Past revolutionary discoveries tended to happen by serendipity, finding something in material families outside of what was known at the time. Our approach, relying only on stoichiometry and a measure of "distance" from what is currently known, is more likely to find novel materials of interest and a sense of where unexplored but promising materials lie compared to ML-guided approaches that proceed within only a given family of materials.

This approach improves performance with experience, in that with every closing of the loop, the ML model undergoes feedback and refinement, enabling efficient exploration of materials space. These improvements ultimately will reduce the cost of materials development and discovery. The success of this approach has been demonstrated by discoveries and rediscoveries coming from vastly different families, illustrating the potential of this tool for the discovery of novel materials with targeted properties. This methodology can be expanded to target more than one desired property, and applied to domains beyond superconductors as long as a mechanism for new data acquisition based on ML-based predictions can be leveraged.

## Materials and Methods

**A. Data.** Our general data source containing the superconducting transition temperature, $T_c$ of many known compounds is the SuperCon database (22), published by the Japanese National Institute for Materials Science. More details are available in the SI.

In this work, we use the version of SuperCon released by Stanev *et al.* (11), available online. This contains 16,414 material compositions and associated critical temperature measurements. However, some of these compositions are invalid (e.g., `Y2C2Br0.5!1.5`) and were removed prior to analysis. Our final training dataset has 16,304 valid compositions. In the Extended Data and the SI, we give additional detail about our training dataset. Figure S1 shows the distribution of $T_c$ values in our training data—note that the distribution is weighted toward low-$T_c$ compositions.

We use MP (23) and OQMD (24) as the set of candidates to screen with ML for superconducting potential. MP and OQMD are some of the largest public sets of computational materials data. Their records contain full crystallographic information for material structures, along with some associated electronic and mechanical properties (but not, importantly, $T_c$). We scraped MP for material records present in it as of October 2020 using the `MPRester` class from the `pymatgen` (51) package. We later downloaded the entire OQMD v1.4 database. The Extended Data contains a table of MP and OQMD material IDs used in this study.

**B. Computational Methods and Uncertainty.** RooSt (52) is a graph neural network (53) that relates material composition to properties by applying a message-passing scheme (54) to a weighted graph representation of the composition's stoichiometry, producing a real-valued embedding vector. To make a prediction, this embedding is then passed through a feedforward network.

In this work, we make use of the publicly-available implementation of RooSt, which is implemented in PyTorch (55). Furthermore, we use the default hyperparameters recommended by the RooSt authors, including basing the initial species representation vectors on the *matscholar* embedding (56). Since we seek materials likely to be high-$T_c$ superconductors, and we expect RooSt's classification model to poorly generalize on out of distribution data, we filter for materials predicted to be in the highest $T_c$ tertile ($T_c \geq 20$ K) with a classification score of at least 0.66 (see SI).

RooSt models incorporate two sources of uncertainty in their $T_c$ predictions: We account for aleatoric uncertainty (randomness of input data) by letting a model estimate a mean and standard deviation for each label's logit (57), and we incorporate epistemic uncertainty (error in the model's result, itself) by averaging over an ensemble of independently trained RooSt models (58).

**C. Experiment.** To synthesize compounds in a medium-throughput manner, arc melting and solid state techniques were used. The standard sample size was 500-700 mg. A list of precursors used in this project is found in Table S3 in the SI and details of the synthetic procedures are found in the SI. Additional heat treatments were performed on an as-needed basis when isolating superconducting phases.

Powder X-ray diffraction patterns were collected at room temperature on the as-melted samples using a Bruker D8 Focus powder diffractometer with Cu-K$\alpha$ radiation ($\lambda_{k,\alpha,1}$ =1.540 596 Å, $\lambda_{k,\alpha,2}$ =1.544 493 Å), Soller slits, and a LynxEye detector to verify the presence of the target phase. We measured from $2\theta$=5°-60° with a step size of 0.018 563° over 4 minutes as an initial screen. When gathering XRD patterns of samples in preparation for Rietveld refinement, 4 h measurements were performed from $2\theta$=5°-120° with a step size of 0.017 15°.

AC-susceptibility measurements were conducted using either a Quantum Design Magnetic Properties Measurement (MPMS) System ($H_{DC}$ =10 Oe, $H_{AC}$=1-3 Oe, 900 Hz) or a Quantum Design Physical Properties Measurement (PPMS) System ($H_{DC}$ =10 Oe,

$H_{AC} =3$ Oe, 1 kHz), measuring T $\geq 2$ K. Since prior density function theory (DFT) calculations (59) suggested that CaAg$_2$Ge$_2$ would superconduct near $T = 1.5$ K, we used the $^3$He option with the MPMS to measure from 0.4 K to 1.7 K for that sample in addition to our standard measurement above 2 K.

**ACKNOWLEDGMENTS.** The authors gratefully acknowledge internal financial support from the Johns Hopkins University Applied Physics Laboratory's Independent Research & Development (IR&D) Program for funding portions of this work. The MPMS3 system used for magnetic characterization was funded by the National Science Foundation, Division of Materials Research, Major Research Instrumentation Program, under Grant #1828490.

**Supporting Information Text**

**A. Material descriptors.** Superconducting properties of materials depend in a complex manner on their composition and structure. In this work, we make a simplifying assumption that the superconducting critical temperature of a material, $T_c$, can be estimated primarily based on its composition alone. Microstructure affects $T_c$ but can be treated as a second-order effect, and is thus beyond the present scope. This assumption is commonly-made in other ML-based superconductor work (e.g., (1–3)). On the other hand, crystal structure is essential to determining superconducting properties and is not trivially derived from composition. In principle, however, stable crystal structures can be implicitly associated with each composition given thermodynamic conditions. In attempting to map composition directly to $T_c$, we therefore test a hypothesis that models can be trained to make those implicit associations with crystal structure as needed, given only information about composition and properties. This continues in the direction of prior work using both vector-based (2, 4, 5) and graph-based (6) representations of materials based only on composition to predict properties.

A critical challenge is therefore to implement material representations and model architectures that balance expressiveness and efficiency. That is, they should capture sufficiently complex physical information to learn subtle effects of composition on $T_c$, but they should not be so complex to require larger amounts of training data than practically obtainable.

**B. Problem Setup and Model Validation.** We formulate our prediction problem as an uncertainty-aware classification task. As shown in Figure S1, the distribution of $T_c$ values in Supercon is skewed, with a large number of materials having $T_c$s close to 0 K. We choose to discretize $T_c$ into three categories, based roughly on tertiles: materials with a measured $T_c$ less than 2 K, materials with a $T_c$ between 2 K and 20 K, and materials with a $T_c$ above 20 K.

Our Representation learning from Stoichiometry (RooSt) models (6) incorporate two sources of uncertainty in their $T_c$ predictions: We account for aleatoric uncertainty by letting a model estimate a mean and standard deviation for each label's logit (7), and we incorporate epistemic uncertainty by averaging over an ensemble of independently trained RooSt models (8).

SuperCon provides data as a validation experiment for our model – can RooSt successfully predict the $T_c$ tertile of unknown materials? We evaluate this question in two settings; the first under a standard uniform cross-validation (Uniform-CV) split of SuperCon, and the second with the leave-one-cluster-out cross validation (leave-one-cluster-out cross-validation (LOCO-CV)) strategy (9). In this approach, we apply $K$-means clustering to the Magpie (4) representation of SuperCon and then train $K$ RooSt models, iteratively holding out each cluster as a test set. Since the clustering will put materials that are similar to each other in the same cluster, LOCO-CV is a better proxy for assessing how well our model will perform when used to identify superconductor candidates in Materials Project.

In this study, we set $K = 3$ for the clustering and summarize cluster characteristics in Table S2 and Figure S2. Note that even this simple clustering procedure has produced inter-cluster heterogeneity – e.g., Cluster 0 is significantly smaller than the other clusters, and Cluster 1 has the bulk of the $20 \leq T_c$ superconductors.

In Figures S3 and S4, we show the results of our study. In the Uniform-CV setting, our model does well – it shows little evidence of overfitting and performs well for all three $T_c$ categories. However, in LOCO-CV, performance degrades significantly and is also much more variable, based on what cluster is being used as the test set. Our result here echoes (2), who show that models trained only on iron-based superconductors fail to accurately predict properties of cuprates, and vice versa.

These results indicate that we should not expect an ML model trained only on SuperCon to consistently identify superconductors in out-of-distribution data, and, as points in SuperCon are more similar to each other than points in Materials Project (Figure 1 of the main text), the LOCO-CV results here are optimistic compared to our actual problem of interest. This motivates our need for multiple iterations of model training, candidate selection, candidate synthesis, and model retraining.

**C. Message-passing over stoichiometry graphs.** Here we go into more detail about the approach used in RooSt (6) to map a material composition to a predicted property. The composition is encoded as a fully-connected graph, where each node $i$ corresponds to an atomic species found in that composition. Each species $i$ has an initial representation vector $h_i^0$ specified by the user (e.g., a one-hot encoding of species type), and the fraction $w_i$ of the composition it makes up (e.g., for $Zr_3Fe_4Sn_4$, $w_{Zr} = 3/15$). A series of $t = 0, \ldots, T-1$ message-passing updates iteratively update each species representation vector to $h_i^{t+1}$ as follows:

$$\begin{align}
e_{i,j}^{t,m} &= f^{t,m}(h_i^t, h_j^t) \\
a_{i,j}^{t,m} &= \frac{w_j \exp(e_{i,j}^{t,m})}{\sum_k w_k \exp(e_{i,k}^{t,m})} \\
h_i^{t+1} &= h_i^t + \sum_{m,j} a_{i,j}^{t,m} g^{t,m}(h_i^t, h_j^t),
\end{align}$$

where $i$ and $j$ are species indices, $f^{t,m}$ and $g^{t,m}$ are multi-layer perceptrons, $a_{i,j}^{t,m}$ are soft attention coefficients computed with a weighted softmax, and the index $m$ denotes separate "attention heads" that stabilize training and improve model expressivity.

The final species representation vectors $h_i^T$ are gathered into a material-level representation vector $h$ after a final soft-attention

operation and then fed through a final MLP $f^y$ to yield a predicted property $\hat{y}$:

$$\begin{aligned}
e_i^{T,m} &= f^{T,m}(h_i^T) \\
a_i^{T,m} &= \frac{w_i \exp(e_i^{T,m})}{\sum_j w_j \exp(e_j^{T,m})} \\
h &= \sum_{m,i} a_i^{T,m} g^{T,m}(h_i^T) \\
\hat{y} &= f^y(h)
\end{aligned}$$

Note that pure (unary) elemental materials are not amenable to this representation and cannot be ingested by a RooSt model.

All of the model's weights for the message-passing and prediction mechanisms (i.e., each $f^{t,m}$ and $g^{t,m}$, and $f^y$, are trained in an end-to-end manner using a cross-entropy loss.

**D. Synthesis details.** For arc melting, more volatile components were wrapped in metallic foils of less-volatile components and melted with a full-scale current of ∼50-100 A with ∼0.7 atm argon present in the vacuum chamber. Zirconium metal, which reacts quickly with oxygen, was first melted in the chamber after it was filled with argon to remove any remaining oxygen before the samples were melted. Samples were melted 2-3 times, flipped, and then melted 2-3 times. If evidence of a superconducting transition was observed, homogenizing heat treatments were performed at temperatures that depended on the chemistry of the sample (generally near 1000 °C for times ranging from overnight to several days).

For metals containing magnesium or calcium, due to their tendency to vaporize, arc melting was not feasible. $CaIn_6Cu_6$, $Ca(Al_2Cu)_4$, $Ca(AgGe)_2$, $Ca_2Zn_3Ag$, $Ca_3(Cu_2Sn)_4$, and $CaCu_9Sn_4$ were made by wrapping samples in tantalum foil (to minimize reactions with the quartz ampoule) and sealing them in evacuated quartz ampoules. The samples were then heated to 700 °C for 2 h and the furnace was turned off and allowed to naturally cool. The same was done for $Ca_3(Cu_2Sn)_4$, $CaCu_9Sn_4$, $Ca(CuSi_{0.33}Sn_{0.33}Ge_{0.33})_2$, $Ca_2Zn_3Ag$, and $Ca(AgGe)_2$ except the anneal was performed at 900 °C. Vaporization of magnesium in ampoules was more of an issue for the Mg-containing samples compared to calcium vaporization in the Ca-containing samples. Samples with nominal MgCuSn, and MgCuBi compositions were similarly wrapped in tantalum foil and sealed in evacuated quartz ampoules. The samples were then heated in a box furnace to 900 °C for 2 h. The furnace was then turned off and the samples allowed to cool to room-temperature. Since tantalum foil has a $T_c$ of 4 K, it was important to separate the samples from the foil before measuring AC susceptibility. This was much more difficult for some of the Mg-containing samples so we also attempted to make MgCuSn and $Mg_2Cu_3Si$ wrapped in molybdenum foil using the same procedures. These samples tended to react with the molybdenum foil and, therefore, did not form the targeted phases.

**E. Isolating the new Zr-In-Ni superconductor.** The superconducting signal was observed both in single crystals and arc melted samples with indium-rich compositions containing $ZrNi_2In$, indium, and $Zr_2Ni_2In$, although the strength of this transition varied throughout a given sample and between samples with the same nominal composition. Nominally $Zr_2Ni_2In$ samples confirmed to contain $Zr_2Ni_2In$ did not superconduct at all. The $T_c$ of indium is well-known and cannot explain the ∼9 K signal. As the strength of the applied static magnetic field was increased to suppress superconductivity in indium, the higher-temperature transition remained and shifted to around 7.1 K. This is what one would expect for a major superconducting phase. Further work is necessary to clarify the precise phase responsible for the superconducting signal. It is clear that, if it is a $ZrNi_2In$-related phase contributing to the signal, indium-rich stoichiometries are necessary for there to be a significant fraction of the superconducting phase since the signal was weak in relatively phase-pure $ZrNi_2In$ and gained strength under more indium-rich conditions.

Insights into the source of the ∼9 K transition can be gained by examining the broader Zr-Ni-Al-Ga-In system. A similar 8-9 K phase transition has been observed in mixed $ZrNi_2Al_xGa_{1-x}$ alloys (10). The $T_c$'s of $ZrNi_2Al$ and $ZrNi_2Ga$ in that work matched prior reports ($T_c$=∼3 K for $ZrNi_2Ga$, $T_c$=∼2 K for $ZrNi_2Al$), demonstrating that the elevated $T_c$ is related to Al-Ga alloying. Such behavior was only observed for substitution on the column-13 site (10). A clear increase in the 8-9 K signal was observed when the Al/Ga ratio decreased from 1 to 0.25. This suggests that Heusler alloys are capable of superconducting with onset temperatures near 8-9 K. Furthermore, this observed behavior is not caused by random impurities or air leaks. Due to the sensitivity of the volume fraction of the new Zr-In-Ni superconductor reported here to different processing conditions, it is likely that the responsible phase is highly sensitive to order or disorder and is easy to miss without a thorough analysis.

One alternate explanation of the ∼9 K signal that we investigated in detail was the possibility of ZrN traces contributing to the signal. Gold-colored flecks were observed in some of the strongly superconducting arc melted samples and cubic ZrN is the only known gold-colored phase that had any chance of being present. Cubic ZrN has a $T_c$ of 9.8-10.0 K, so an air leak might explain the observed ∼9 K $T_c$ (11). Samples shown by XRD to definitively contain ZrN ($Zr_2InNi$ and ZrNi, melted under nitrogen) showed a $T_c$ below 6 K. No superconducting transition was observed in ZrNi samples melted under argon, demonstrating that indium is necessary for the ∼9 K $T_c$.

The strongest signal that we achieved in arc melted samples came from a sample melted under argon and annealed for 5.5 days at 1000 °C. This anneal allowed indium to vapor transport away from the bulk of the sample. Parts of the sample with a shiny gray and gold appearance had a comparatively strong superconducting signal (Figure S6a).

.

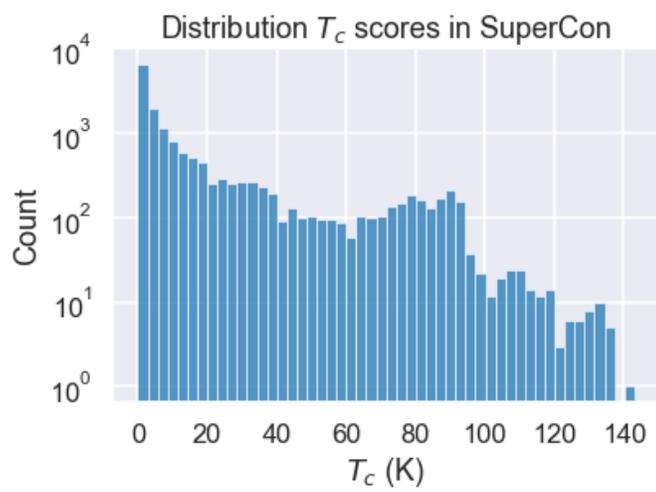

**Fig. S1.** Distribution of raw $T_c$ values in the version of SuperCon used as training data. The data are long-tailed with a small number of high-$T_c$ outliers.

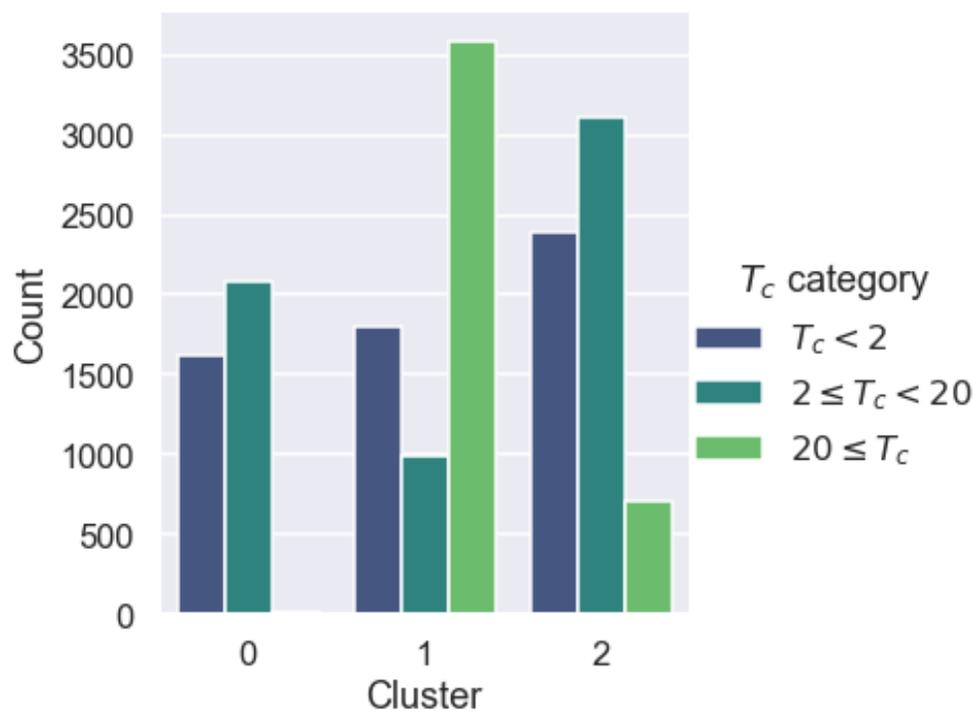

**Fig. S2.** Statistics of $T_c$ across clusters used in LOCO-CV study

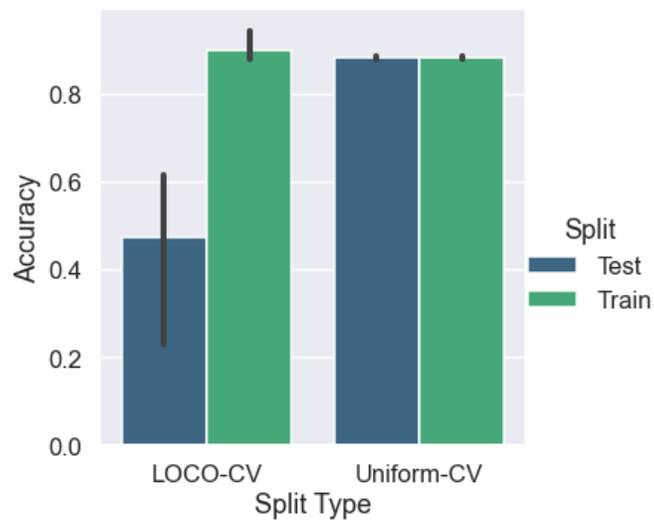

**Fig. S3.** Training and test set accuracies for Uniform-CV vs. LOCO-CV, averaged over each fold and clusters. Bars show 95% confidence intervals for the standard error of the mean estimate. The model severely overfits in the LOCO-CV case, and its test set accuracy is much more cluster-dependent and variable.

|              | $T_c < 2$ | $2 \leq T_c < 20$ | $20 \leq T_c$ |
|--------------|-----------|-------------------|---------------|
| # Materials  | 5,802     | 6,182             | 4,320         |

**Table S1. Distribution of $T_c$ values by tertile in our classification formulation. We treat the prediction problem as a three-class classification problem.**

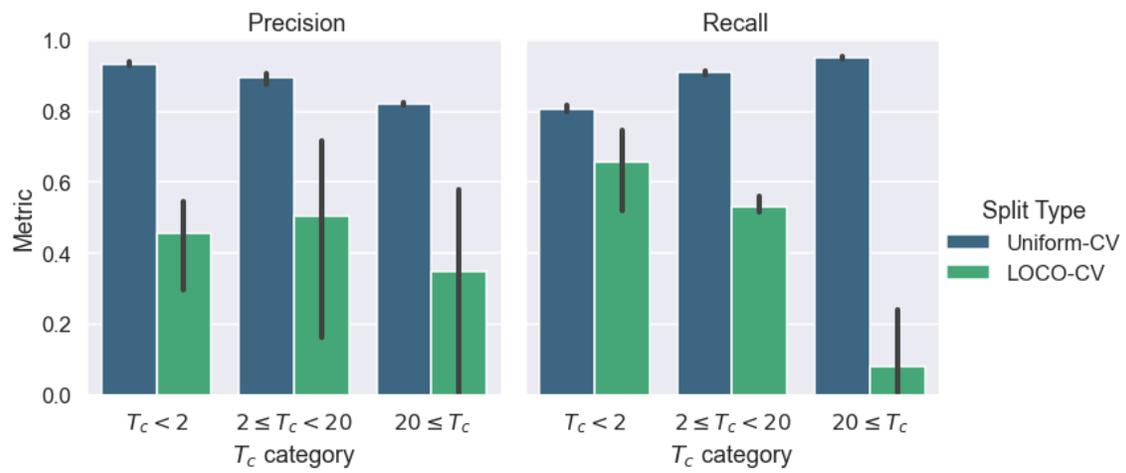

**Fig. S4.** Test set precision and recall analysis for each $T_c$ category for the uniform vs. LOCO-CV study, averaged over each fold and cluster. Bars show 95% confidence intervals for the standard error of the mean estimate. The model's metrics are much more variable and cluster-dependent for the LOCO-CV model.

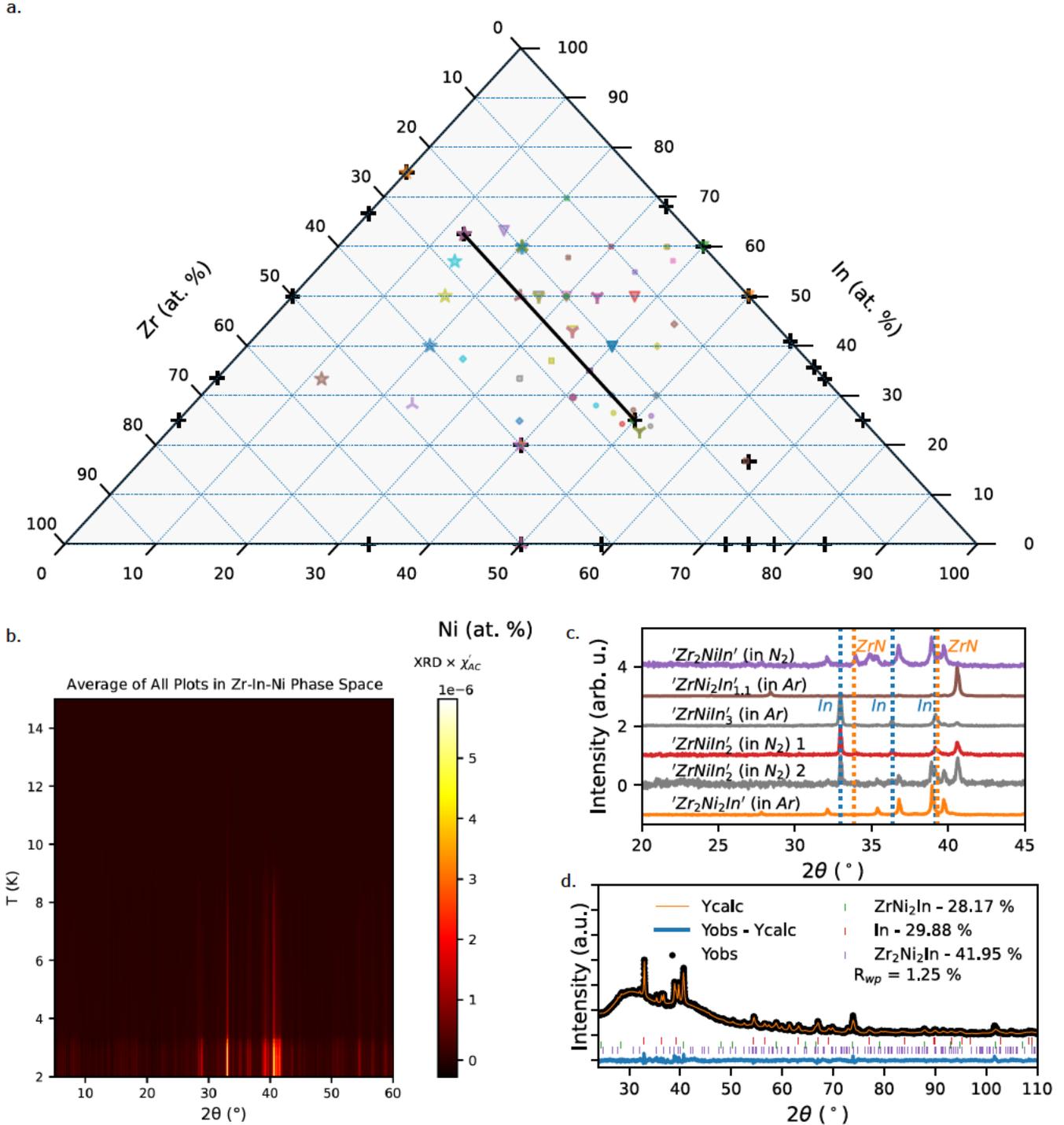

**Fig. S5.** Samples from around the Zr-In-Ni phase diagram (a) were made, including several binary alloys. XRD data was used with the $\chi'_{AC}$ data to identify the new superconductor in the Zr-In-Ni system. (b) shows a heat map averaged over all the samples in the system where the color equals the normalized XRD intensity multiplied by $\chi'_{AC}$ ($color = -\chi'_{AC} \times Intensity_{XRD}$). Indium is consistently present in samples with strong superconducting signals below 3.4 K but cannot explain the ~9 K transition that is consistently observed. (c) and (d) show how XRD was used to identify the phases present in samples with and without strong superconducting signals. Indium diffraction peaks were used as an internal standard to correct for sample height variations since In-rich samples were extremely malleable and could not be made into a powder. The nominally $ZrNi_2In_{1.1}$ sample contained only indium and $ZrNi_2In$. The nominally $Zr_2Ni_2In$ sample was primarily $Zr_2Ni_2In$. Reitveld refinements of samples containing a strong superconducting signal (d) showed indium, $ZrNi_2In$, and $Zr_2Ni_2In$ present. Our $Zr_2Ni_2In$ sample did not superconduct, implying that $ZrNi_2In$ does superconduct under certain indium-rich conditions. To rule out ZrN impurities as the superconducting phase, a number of samples were melted in nitrogen. Samples intentionally made with the largest amount of ZrN present based on XRD had a $T_c$ below 6 K and, therefore, ZrN impurities would not explain the ~9 K transition.

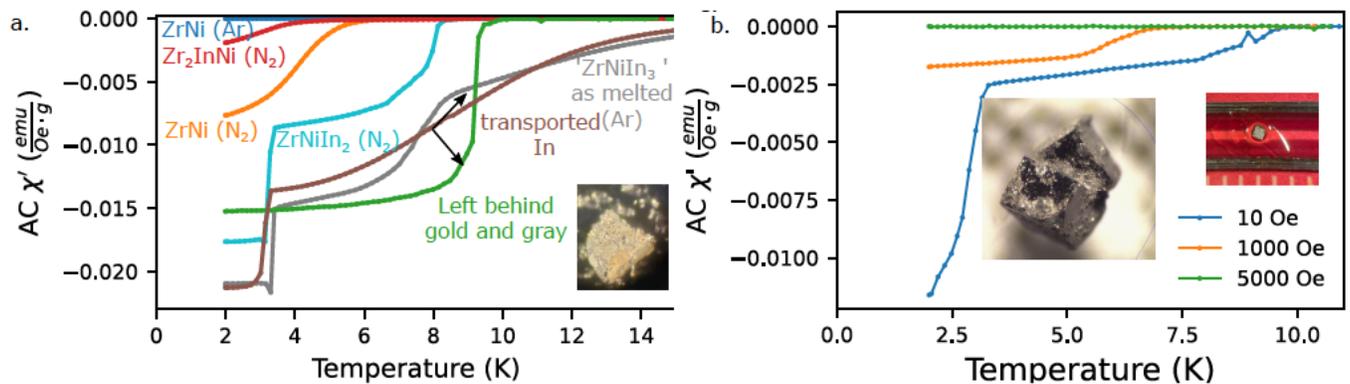

Fig. S6. The existence of the ∼9 K signal was robust for both arc melted samples and single crystals. To rule out ZrN impurities as the source of the superconducting signal, we arc melted samples of ZrNi and $Zr_2NiIn$, and $ZrNiIn_2$ stoichiometry under nitrogen (a). ZrN was identified using XRD in these samples. When ZrNi was melted under argon like most of our samples containing the ∼9 K signal, no superconducting transition was observed. The ∼9 K signal was only visible in the sample with a nominal $ZrNiIn_2$ composition, indicating that indium-rich compositions are required for the ∼9 K transition to appear. Indium acted as a useful internal standard such that the strength of its superconducting signal could be compared to the strength of the ∼9 K signal to determine the relative fraction of sample that superconducts at different temperatures. The strength of the indium signal is relatively comparable to the strength of the ∼9 K signal (when indium could be clearly observed using XRD), indicating that the ∼9 K transition is a bulk behavior. Furthermore, when indium was vapor transported away from the bulk of an indium-rich sample arc melted under argon, the leftover bulk of material exhibited a strong superconducting signal without indium. A photo of this material is shown in (a). Single crystals grown in an indium flux (b) contained varying amounts of the ∼9 K superconducting phase, demonstrating that some amount of the ∼9 K superconductor forms in materials made using vastly different growth techniques. As the magnetic field increased, the ∼9 K signal shifted to lower temperatures and, by 5000 Oe, disappeared. This is further evidence that the ∼9 K transition is a bulk behavior.

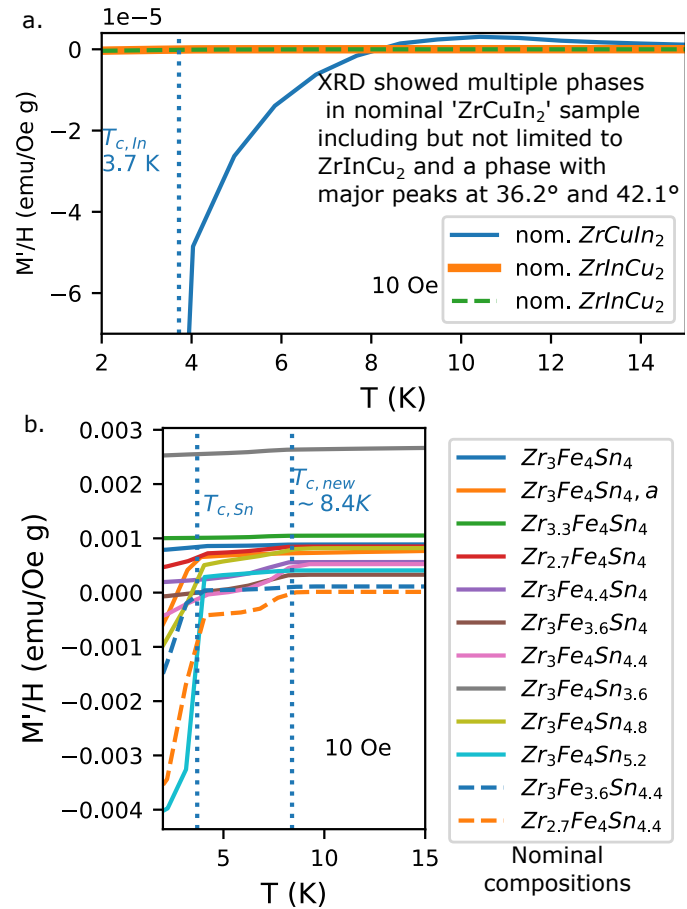

**Fig. S7.** We saw similar anomalous signals in nominally ZrCuIn$_2$ and Zr$_3$Fe$_4$Sn$_4$ samples but the superconducting signals also proved difficult to isolate.

| Cluster | Size |
|---:|---:|
| 0 | 3,723 |
| 1 | 6,377 |
| 2 | 6,204 |

**Table S2. Sizes of clusters used in LOCO-CV study**

**Table S3. Precursor List**

| Element or Compound | Source | Form | Purity |
|---|---|---|---|
| Mg | Alfa Aesar | turnings, 3.2 mm wide | 99.8% |
| MgSi$_2$ | Beantown Chemical | powder | 99.5% |
| Ca | Alfa Aesar | granules, -16 mesh | 99.5 % |
| Sc | Alfa Aesar | pieces | 99.9 % |
| Ti | Alfa Aesar | foil, $0.25\,\mathrm{mm}$ | 99.5 % |
| Zr | Strem | foil | 99.8% |
| Hf | Alfa Aesar | foil, 0.025 mm | 99.5 % |
| V | Alfa Aesar | foil, $0.127\,\mathrm{mm}$ | 99.8 % |
|  | Alfa Aesar | wire, $1.5\,\mathrm{mm}$ | 99.8 % |
| Nb | Alfa Aesar | foil, $0.3\,\mathrm{mm}$ | 99.8 % |
|  | Alfa Aesar | foil, $0.025\,\mathrm{mm}$ | 99.8 % |
|  | Strem | foil | 99.8 % |
| Ta | Alfa Aesar | foil, $0.025\,\mathrm{mm}$ | 99.95 % |
|  | Beantown Chemical | foil, $0.025\,\mathrm{mm}$ | 99.95 % |
| Mo | Beantown Chemical | foil, $0.025\,\mathrm{mm}$ | 99.95 % |
| W | Alfa Aesar | wire, $0.75\,\mathrm{mm}$ diameter | 99.95 % |
| Mn | Beantown Chemical | pieces | 99.95 % |
| Fe | Goodfellow | foil | 99.8 % |
| Co | Alfa Aesar | pieces | 99.9+ % |
| Co | Aldrich | pieces | 99.5 % |
| Ni | Alfa Aesar | foil | 99+% |
|  | Alfa Aesar | -300 mesh | 99.8% |
| Cu | Strem | foil | 99.9 % |
|  | Alfa Aesar | powder, -325 mesh | 99% |
|  | Alfa Aesar | shot, 4-6 mm, Pufratronic | 99.999% |
| Ag | Alfa Aesar | wire | 99.9 % |
| Zn | Alfa Aesar | ingot | 99.99 % |
| B | Alfa Aesar | powder, -325 mesh | 98 % |
| Al | Beantown Chemical | wire, 1.0 mm diameter | 99.999 % |
| Ga | NOAH Tech | chunks | 99.99 % |
| In | Alfa Aesar | shot, 4 mm tear drop | 99.99 % |
| C | Alfa Aesar | graphite rod, 6.15 mm diameter | 99.9995% |
| Si | Alfa Aesar | lump | 99.999+% |
| Ge | Beantown Chemical | pieces, $3\text{-}9\,\mathrm{mm}$ | 99.999% |
| Sn | Beantown Chemical | shot (8-20 mesh) | 99.5 % |
|  | Beantown Chemical | granules | 99.9 % |
| Pb | Alfa Aesar | pieces | 99.999 % |
| P | Sigma Aldrich | red, powder | 97+% |
| Bi | Alfa Aesar | needles, $1\text{-}5\,\mathrm{cm}$, $1.5\,\mathrm{mm}$ diameter | 99.99% |
| S | Alfa Aesar | powder, Puratronic | 99.999% |
| Yb | Alfa Aesar | pieces, sublimed | 99.9 % |